\documentclass[aip,apl,reprint,superscriptaddress%,groupedaddress
]{revtex4-1}
\pdfoutput=1
\usepackage{graphicx}
\usepackage{amsmath,amsfonts,amssymb}
\usepackage{xcolor}
\usepackage{epstopdf}
\renewcommand{\epsilon}{\varepsilon}
\newcommand{\blue}[1]{\textcolor{black}{#1}} %added by authors

\draft % marks overfull lines with a black rule on the right

\begin{document}

%\preprint{}

\title{Water adsorption enhances electrical conductivity in transparent p-type CuI}



\author{Andrea Crovetto}
\email[]{Electronic mail: andrea.crovetto@helmholtz-berlin.de}
%\thanks{}
\affiliation{Department of Structure and Dynamics of Energy Materials, Helmholtz-Zentrum Berlin f\"ur Materialien und Energie, Hahn-Meitner Platz 1, 14109 Berlin, Germany}

\author{Hannes Hempel}
%\thanks{}
\affiliation{Department of Structure and Dynamics of Energy Materials, Helmholtz-Zentrum Berlin f\"ur Materialien und Energie, Hahn-Meitner Platz 1, 14109 Berlin, Germany}

\author{Marin Rusu}
%\thanks{}
\affiliation{Department of Structure and Dynamics of Energy Materials, Helmholtz-Zentrum Berlin f\"ur Materialien und Energie, Hahn-Meitner Platz 1, 14109 Berlin, Germany}

\author{Leo Choubrac}
%\thanks{}
\affiliation{Department of Structure and Dynamics of Energy Materials, Helmholtz-Zentrum Berlin f\"ur Materialien und Energie, Hahn-Meitner Platz 1, 14109 Berlin, Germany}

\author{Danny Kojda}
%\thanks{}
\affiliation{Department of Methods for Characterization of Transport Phenomena in Energy Materials, Helmholtz-Zentrum Berlin f\"ur Materialien und Energie, Hahn-Meitner Platz 1, 14109 Berlin, Germany}

\author{Klaus Habicht}
%\thanks{}
\affiliation{Department of Methods for Characterization of Transport Phenomena in Energy Materials, Helmholtz-Zentrum Berlin f\"ur Materialien und Energie, Hahn-Meitner Platz 1, 14109 Berlin, Germany}
\affiliation{\blue{Institute of Physics and Astronomy, University of Potsdam, Karl-Liebknecht-Str. 24-25, D-14476 Potsdam}}

\author{Thomas Unold}
%\thanks{}
\affiliation{Department of Structure and Dynamics of Energy Materials, Helmholtz-Zentrum Berlin f\"ur Materialien und Energie, Hahn-Meitner Platz 1, 14109 Berlin, Germany}

%\date{\today}

\begin{abstract}
CuI has been recently rediscovered as a p-type transparent conductor with a high figure of merit. Even though many metal iodides are hygroscopic, the effect of moisture on the electrical properties of CuI has not been clarified.
%Unlike most n-type transparent conductive oxides,
In this work, we \blue{observe} a two-fold increase in the conductivity of CuI after exposure to ambient humidity for 5 hours, followed by slight long-term degradation. Simultaneously, the work function of CuI decreases by almost 1~eV, which can explain the large spread in the previously reported work function values. The conductivity increase is partially reversible and is maximized at intermediate humidity levels. \blue{Based on the large intra-grain mobility measured by THz spectroscopy, we suggest that hydration of grain boundaries may be beneficial for the overall hole mobility}.

\end{abstract}


\maketitle %\maketitle must follow title, authors, abstract and \pacs


\section{Introduction}
Electron-doped transparent conductive materials (n-type TCMs) are used routinely as optically transparent electrical contacts in solar cells and displays.~\cite{Ellmer2012,Morales-Masis2017,Gordon2000,Crovetto2016a} The performance of a given thin-film material as a TCM can be quantified using one of the several figures of merit (FOM) proposed by various authors.~\cite{Ellmer2012,Morales-Masis2017,Gordon2000} A widely adopted FOM is simply the ratio between the electrical conductivity and the absorption coefficient of the material averaged over the visible region.~\cite{Gordon2000} A few different n-type oxides have FOMs between 1~$\Omega^{-1}$ and 10~$\Omega^{-1}$, which are sufficiently high to satisfy the requirements of many optoelectronic devices.~\cite{Gordon2000} Conversely, the development of hole-doped (p-type) TCMs has proven more difficult \blue{due to the deep valence band and high hole effective masses found in many wide band-gap p-type semiconductors.~\cite{Morales-Masis2017,Fioretti2020,Kawazoe1997}
As a consequence,} the FOM of the current state-of-the-art p-type TCMs is more than two orders of magnitude lower than in the best n-type oxides.~\cite{Yang2016a,Woods-Robinson2018} \blue{If this limitation were overcome, the design options for optoelectronic devices would be greatly expanded and fully transparent electronics could be realized.~\cite{Morales-Masis2017,Kawazoe1997,Fioretti2020}}

A promising p-type TCM is the binary compound CuI. Although its transparency and conductivity have been known for over a century,~\cite{Badeker1907} it was only recently discovered that growing CuI by reactive sputtering could enhance its FOM up to the highest values observed for any p-type TCMs.~\cite{Yang2016a} Since these results were achieved without any (intentional) extrinsic doping, CuI has significant potential for further improvement by the controlled incorporation of acceptor impurities such as the Group VI elements O, S, and Se.~\cite{Grauzinyte2019} CuI has several electronic features that can promote a high hole conductivity by native doping: an antibonding valence band with strong dispersion due to Cu d-I p hybridization,~\cite{Li2018b} a low formation energy for Cu vacancies,~\cite{Jaschik2019} the lack of compensating native dopants,~\cite{Wang2011b} and the clustering of Cu vacancies into ordered structures that stabilize a Cu-deficient stoichiometry.~\cite{Jaschik2019}

Interestingly, many halide compounds such as SnI$_2$, PbI$_2$, and hybride halide perovksites (CH$_3$NH$_3$PbI$_3$ and similar) undergo major bulk reactions and structural changes when exposed to moisture.\cite{Leguy2015} On the other hand, several works in the perovskite and organic solar cell literature refer to CuI as a hydrophobic and air-stable hole transport layer based on contact angle measurements and its structural stability in air.\cite{Shao2012a,Sun2014,Chen2015} Even though CuI is structurally stable under ambient conditions, properties of crucial importance for TCM applications (e.g., electrical conductivity and work function) can be sensitive to much smaller amounts of adsorbed water or oxygen than the quantities necessary for bulk reactions. Previous work on stability of the electrical conductivity in CuI focused only on long time scales (weeks or months).~\cite{Yang2016a,Raj2019}
%and generally found minimal degradation upon prolonged exposure to ambient air.
In a similar fashion, the CuI work function reported in several papers was measured after an undefined exposure time to air, without investigating the effect of oxygen or moisture on the work function of a pristine CuI surface.~\cite{Rojas2016,Kaushik2017,Das2015,Yoon2016,Shao2012,Sun2014,Jeon2018} In this work, we investigate changes in the conductivity, work function, and ionization \blue{energy} of CuI in the first few hours of exposure to air at different relative humidities (RH).

\section{Experimental details}
\begin{figure}[t!]
\centering%
\includegraphics[width=\columnwidth]{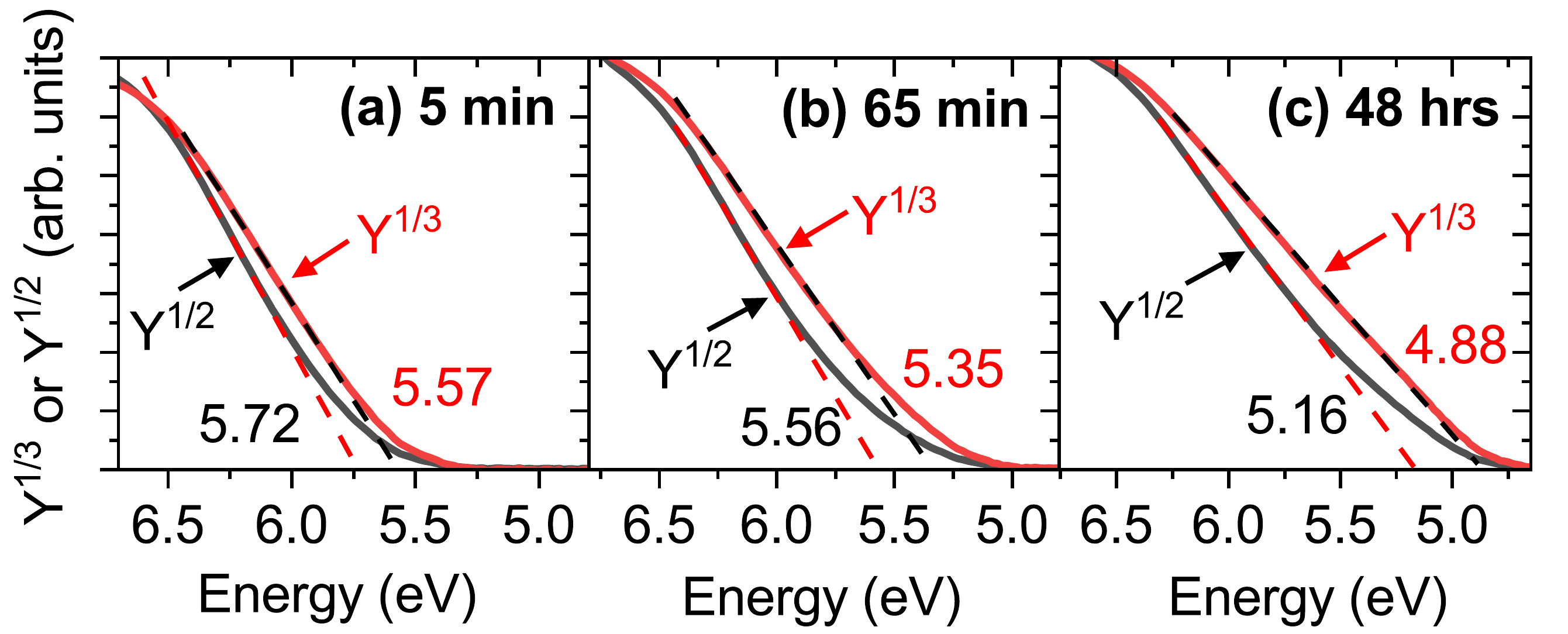}
\caption{Determination of the CuI ionization \blue{energy} by \blue{ambient pressure} photoemission spectroscopy 5 min after CuI film growth (a), 65 min after CuI film growth (b), and 48 h after CuI film growth (c). \blue{Under the assumption of a nondegenerate (degenerate) semiconductor, the ionization energy can be extracted by linear regression and extrapolation of the $Y^{\frac{1}{3}}$ spectrum ($Y^{\frac{1}{2}}$ spectrum), where $Y$ is the photoelectron yield.~\cite{Baikie2014,Harwell2016} The extrapolated values (in eV) are shown. Spectra are normalized.}}
\label{fig:pes}
\end{figure}

250~nm-thick CuI films were grown on fused silica glass by gas-phase iodization of sputter-deposited metallic Cu films, similar to previous reports.~\cite{Badeker1907,Crovetto2020a} Iodization was performed by placing the Cu films on a hotplate next to iodine pellets, covering both the films and a pellet with an upside-down Petri dish, \blue{heating the hotplate to 100$^\circ$C over 2~min}, and removing the iodized films from the hotplate 5~min after reaching the temperature setpoint. The films were immediately placed into a nitrogen-filled transport box \blue{before characterizing either the electrical conductivity or the surface electronic properties. The time between growth interruption and the first measurement point was 5~min for conductivity measurements and 10~min for surface measurements. Conductivity measurements were performed with a collinear four-point probe in a humidity-controlled chamber \blue{under normal indoor lighting}. Work function and ionization \blue{energy} measurements were performed in the dark} using a combined Kelvin probe (KP)/\blue{ambient pressure} photoemission spectroscopy (PES) system (SKP-5050 and APS-02, KP Technology). Work function and ionization \blue{energy} were measured sequentially on the same sample as a function of ambient exposure time by using the same tip as a Kelvin probe and as a photoelectron detector.~\cite{Baikie2014}
The ionization \blue{energy} (Fig.~\ref{fig:pes}) can be determined by \blue{linear regression and extrapolation of $Y^{\frac{1}{3}}(E)$ spectra or of $Y^{\frac{1}{2}}(E)$ spectra as shown in Fig.~\ref{fig:pes}}. $Y(E)$ is the \blue{photoelectron yield versus photon energy $E$, measured by} illuminating the sample with monochromated UV light.~\cite{Baikie2014} \blue{The $Y^{\frac{1}{3}}$ method is appropriate for nondegenerate semiconductors,~\cite{Harwell2016} whereas the $Y^{\frac{1}{2}}$ method is appropriate for degenerate semiconductors or metals.~\cite{Baikie2014} As will be explained later, the Fermi level of the CuI surface changes as a function of ambient exposure time, so both methods are shown in Fig.~\ref{fig:pes}.}
Work function measurements were calibrated by first measuring the work function of a gold reference sample by \blue{PES using the $Y^{\frac{1}{2}}$ method}, and by then measuring the contact potential difference (CPD) between the gold reference and the KP tip to determine the tip's work function. The work function of the CuI films was then \blue{calculated} by adding the measured CuI-tip CPD to the work function of the tip.

\begin{figure}[t!]
\centering%
\includegraphics[width=\columnwidth]{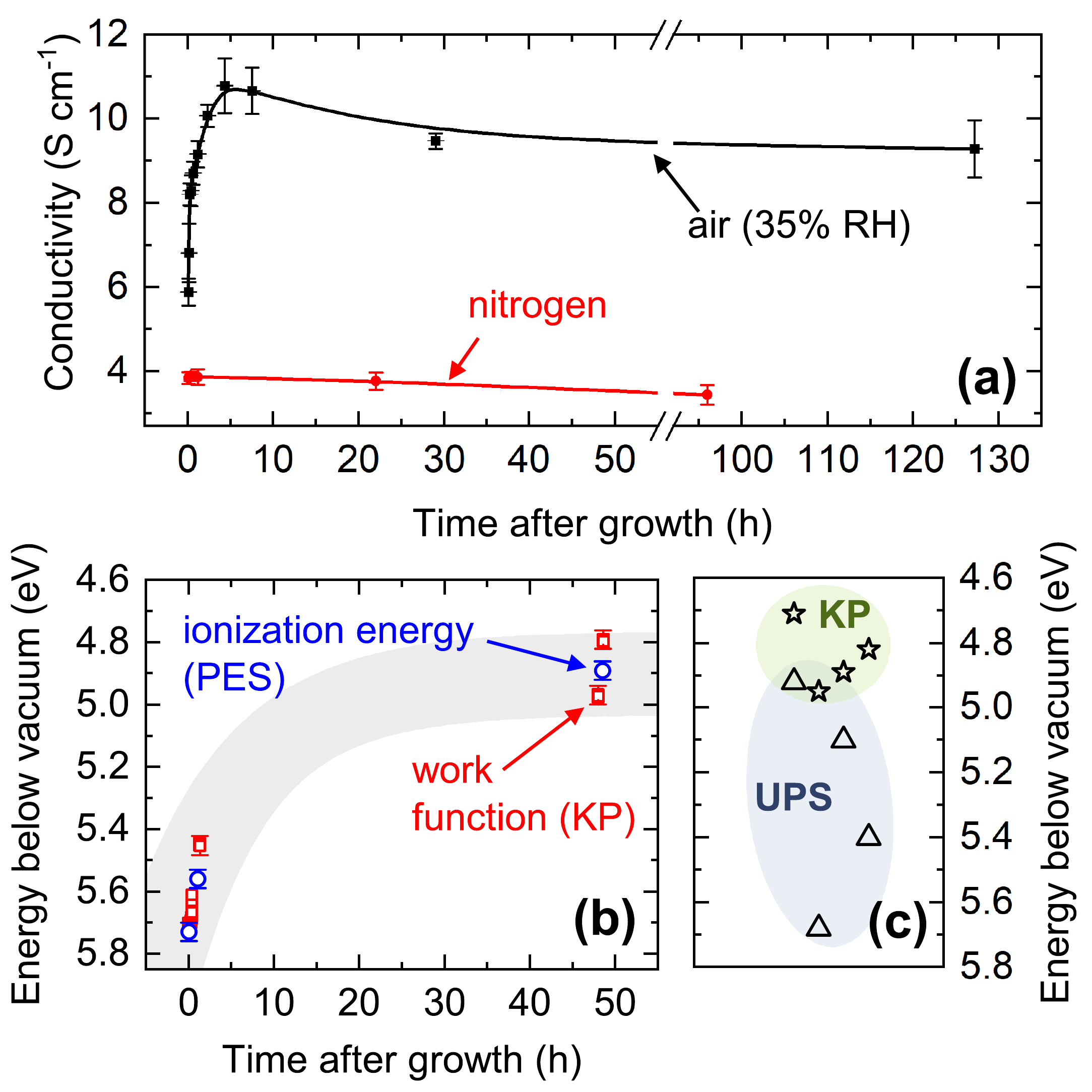}
\caption{(a): Conductivity as a function of time for a CuI film stored in air at 35\%~RH, and for a nominally identical film stored in a glove box with water content below 1~ppm. (b): Work function (as measured by KP) and ionization \blue{energy} (as measured by PES) of a CuI film held at (35$\pm$5)\%~RH. \blue{The shaded region is a guide to the eye. For reasons explained in the main text, the two initial ionization energy data points are determined using the $Y^{\frac{1}{2}}$ method in Fig.~\ref{fig:pes}(a,b), whereas the final data point is determined using the $Y^{\frac{1}{3}}$ method in Fig.~\ref{fig:pes}(c). Error bars are $\pm 30$~meV for both the ionization energy and the work function.} (c): Compilation of work function values measured in previous studies. Star markers are used for KP measurements.~\cite{Rojas2016,Kaushik2017,Das2015} Triangle markers are used for UV photoemission spectroscopy (UPS) measurements.~\cite{Yoon2016,Shao2012,Sun2014,Jeon2018}}
\label{fig:time_dependent}
\end{figure}

\section{Results and discussion}
Basic structural characterization (Figs.~S1,S2, Supporting Information) confirms that the CuI films are in the zincblende phase known for its transparent and conductive properties. Sub-band gap optical transmission is in the 50-80\% range (Fig.~S3, Supporting Information). The time evolution of the conductivity for two nominally identical CuI samples is compared in Fig.~\ref{fig:time_dependent}(a). One sample was kept at 35\% RH in the humidity chamber, the other sample was kept in a nitrogen-filled glove box with oxygen and water content below 1~ppm. The conductivity of the sample stored in an inert atmosphere slightly decreases over two days. Conversely, the sample stored at ambient humidity experiences a conductivity increase by nearly a factor of two over the first 5 hours after growth. The swift conductivity enhancement is followed by some degradation on a much longer time scale. We verified that the time evolution of the conductivity only depends weakly on the type of \blue{background} atmosphere (air or nitrogen) but depends strongly on the presence or absence of moisture.

Another nominally identical sample was employed for work function and ionization \blue{energy} measurements in air with laboratory humidity in the (35$\pm$5)\% RH range. Similarly to the case of the conductivity, work function and ionization \blue{energy} undergo rapid changes in the first few hours after growth (Fig.~\ref{fig:time_dependent}(b)). \blue{After two days of exposure to ambient humidity, the work function has decreased from 5.70~eV to 4.80~eV and the ionization energy has decreased from 5.57~eV to 4.88~eV using the $Y^{\frac{1}{3}}$ method, or from 5.72~eV to 5.16~eV if the $Y^{\frac{1}{2}}$ method is used (Fig.~\ref{fig:pes}).}
%For clarity, we only show the ionization energy measured with the $Y^{\frac{1}{3}}$ method in Fig.~\ref{fig:time_dependent}(b). Both methods are shown in Fig.~S5, Supporting Information.
%Their initial values are 5.70~eV for the work function and 5.57~eV for the ionization \blue{energy} (Fig.~\ref{fig:pes}(a)). After 2 days of exposure to ambient humidity, the work function has decreased to 4.80~eV (0.90~eV lower) and the ionization \blue{energy} to 4.89~eV (0.68~eV lower, Fig.~\ref{fig:pes}(c)).
Work function values reported in the literature~\cite{Rojas2016,Kaushik2017,Das2015,Yoon2016,Shao2012,Sun2014,Jeon2018} are plotted in Fig.~\ref{fig:time_dependent}(c). Comparison to our data suggests that the large spread in the previously measured values may be due to different air exposure times and humidity levels in the previous studies. Consistent with this hypothesis, we note that the reported work functions measured by UV photoemission spectroscopy in ultra-high vacuum are generally higher than those measured by Kelvin probe (Fig.~\ref{fig:time_dependent}(c)), which could be due to (partial) water desorption in a vacuum environment. \blue{The work function change $\Delta \Phi$ is caused by preferential alignment of the dipole moments of adsorbed water molecules, which generates a net surface dipole.} Modeling the water dipole layer as a plane capacitor yields
\begin{equation}
\frac{\Delta \Phi}{e} = \frac{n \mu_{\perp} d}{\varepsilon_0 \varepsilon_\mathrm{r}}
\label{eq:dipole}
\end{equation}

\begin{figure}[t!]
\centering%
\includegraphics[width=\columnwidth]{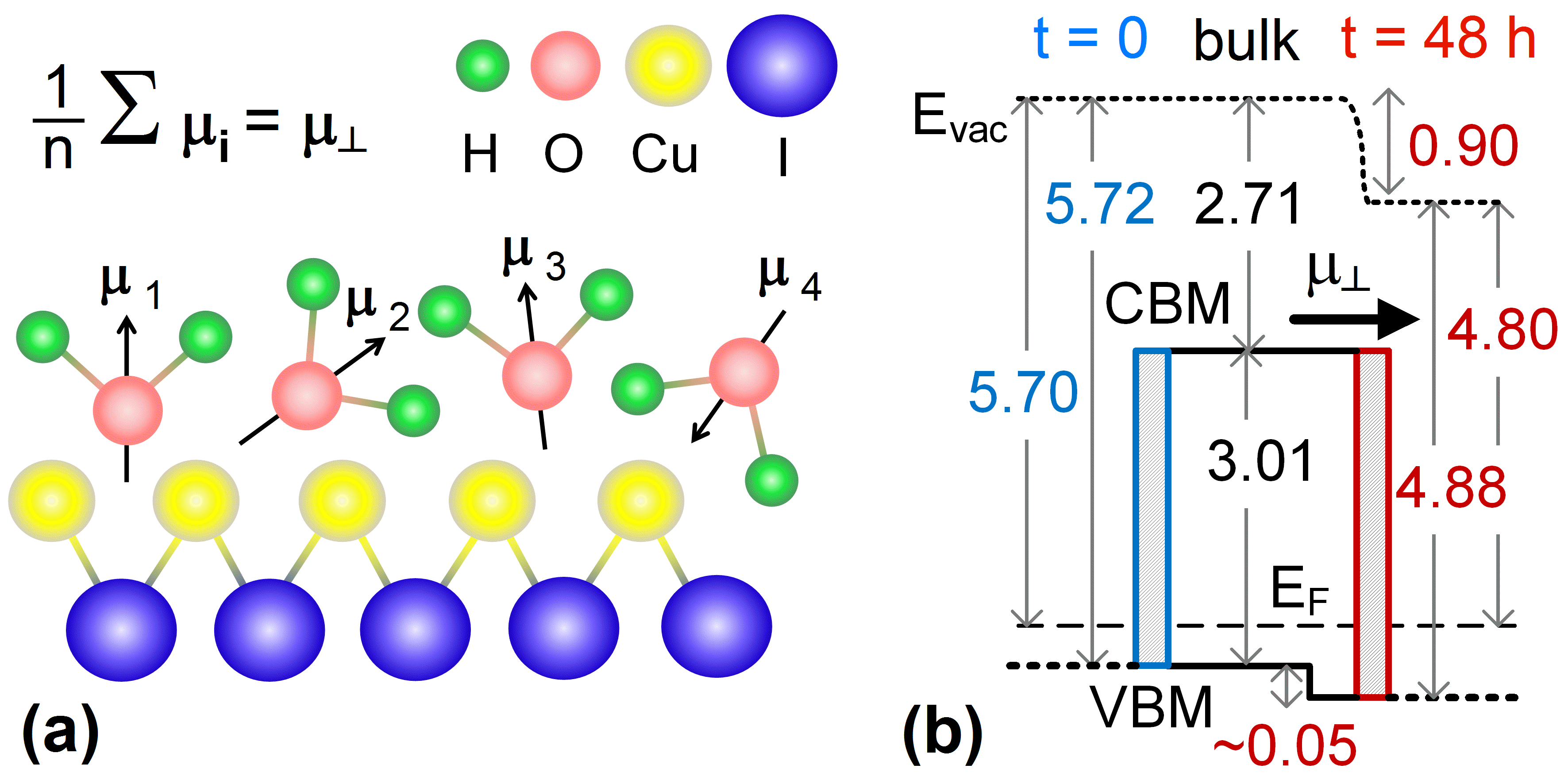}
\caption{(a) Simplified sketch of the mechanism generating a surface dipole by water adsorption \blue{on} CuI. The average perpendicular component of the dipole moment of a water molecule ($\mu_{\perp}$) is given by the vector average of the individual water dipoles $\vec{\mu}_\mathrm{i}$. \blue{The surface is drawn as Cu-terminated due to the high volatility of I.} (b) Band diagram of a CuI surface \blue{shortly after growth ($t = 0$, left side) and after exposure to ambient humidity for 48~hours ($t = 48$~h, right side). VBM and CBM are the valence band maximum and conduction band minimum respectively, and $E_\mathrm{vac}$ is the vacuum level. Ionization energy and work function data are taken from Fig.~\ref{fig:time_dependent}(b).}
%The ionization \blue{energy} is $\mathrm{IE} = \mathrm{VBM} - E_\mathrm{vac} = 5.72$~eV.
%The work function of the moisture-exposed surface is $\Phi = E_\mathrm{F} - E_\mathrm{vac} = 4.80$~eV on the right side of the surface dipole.
The band gap $E_\mathrm{g}$ is estimated as 3.01~eV using a Tauc plot for direct gap materials (Fig.~S4, Supporting Information). ~\blue{Based on this band gap value, the electron affinity of the as-grown surface (2.71~eV) is derived.}}
\label{fig:humidity_model}
\end{figure}

In Eq.~\ref{eq:dipole}, $n \simeq 9.2$~nm$^{-2}$ is the area density of \blue{one monolayer (ML) of adsorbed water molecules,} assuming the equilibrium intermolecular separation of 3.55~\AA~and hexagonal close packing.~\cite{Wensink2000} $\mu_{\perp}$ is the surface-normal component of the vector average of the dipole moments of all water molecules (Fig.~\ref{fig:humidity_model}(a)). $d$ is the number of adsorbed MLs at 35\% RH, which according to the BET theory~\cite{Brunauer1938} is between 0.5~ML (strongly hydrophobic surface) and 1.5~ML (strongly hydrophilic surface). Since contact angle measurements indicate that the CuI surface is an intermediate case,~\cite{Shao2012a,Sun2014} we assume $d =1$~ML. $\varepsilon_0$ and $\varepsilon_\mathrm{r}$ are the permittivities of vacuum and of the water layer respectively. We take $\varepsilon_\mathrm{r} = 1$ because the water molecules generate the dipole field themselves, instead of responding to it as a dielectric. Furthermore, the first ML of adsorbed water is known to be immobile, with $\varepsilon_\mathrm{r} \simeq 1$ measured under an externally applied voltage,~\cite{McCafferty1970} in stark contrast to $\varepsilon_\mathrm{r} \simeq 80$ in bulk water.
Taking $\Delta \Phi = -0.90$~eV as \blue{determined by KP after two days of ambient exposure}, we find that $\mu_{\perp} = 0.14\,\mu_\mathrm{H_2O}$, where $\mu_\mathrm{H_2O}$ is the magnitude of the dipole moment of a single water molecule (1.85~D). Assuming that a close-packed monolayer of adsorbed water has formed on CuI, the surface dipole has therefore around 14\% of the maximum dipole strength of a perfectly aligned water monolayer.
Since the work function decreases upon water adsorption, $\overrightarrow{\mu}_{\perp}$ points away from the CuI film. Therefore adsorbed water is preferentially bonded to the CuI surface through its oxygen atoms (Fig.~\ref{fig:humidity_model}(a)), as systematically observed in many oxides, chalcogenides, and pnictides.~\cite{Stevanovic2014,Mayer1992}

\blue{Two} features of the plot in Fig.~\ref{fig:time_dependent}(b) require further discussion.
The first is the negative shift in the measured work function \blue{just} before and \blue{just} after performing a PES measurement. \blue{The shift is negligible on the as-grown surface but it becomes larger for increasing ambient exposure time. A substantial shift of $-200$~meV is observed 48 hours after growth, as shown in detail in Fig.~S5(c), Supporting Information.}
%The shift is particularly large ($\sim -200$~meV) after the build-up of an adsorbed water layer 50~hours after growth.
This effect can be explained by photoinduced hydrophilicity, a well-known phenomenon in, e.g., various metal oxides and graphene.~\cite{Miyauchi2002,Xu2014} Briefly, the UV radiation used during PES measurements can promote surface hydroxylation by dissociating adsorbed water molecules~\cite{Xu2014} or cause surface structural changes that enhance water adsorption.~\cite{Miyauchi2002} These effects can increase $n$, $d$, or $\mu_{\perp}$ in Eq.~\ref{eq:dipole}, and thus shift the work function to even lower values. However, photoinduced work function changes are at least partially reversible, since the work function slowly increases again with time once UV irradiation is turned off (Fig.~S5(c), Supporting Information).

\begin{figure*}[t!]
\centering%
\includegraphics[width=1\textwidth]{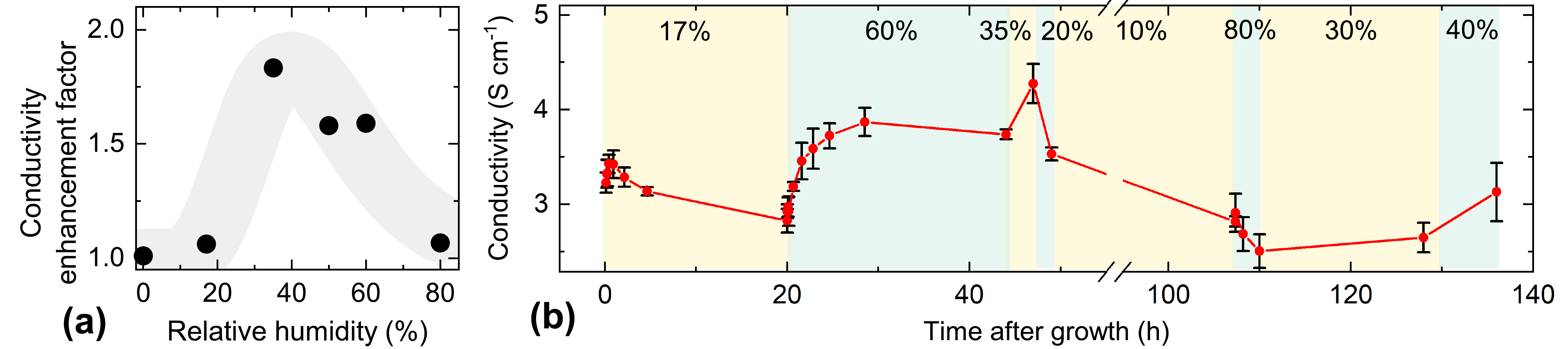}
\caption{(a): Conductivity enhancement factor (maximum conductivity divided by initial conductivity) as a function of relative humidity. Each data point refers to a unique CuI film only exposed to a single humidity. The shaded region is a guide to the eye. (b): Conductivity trace of a single CuI sample sequentially exposed to different humidities.}
\label{fig:various_humidities}
\end{figure*}

\blue{A second issue is the applicability of the $Y^{\frac{1}{3}}$ method versus the $Y^{\frac{1}{2}}$ method to determine the ionization energy of CuI. In the initial stages of water adsorption, the $Y^{\frac{1}{3}}$ method yields an ionization energy that is 130~meV smaller than the work function, implying very high degenerate p-type doping at the surface of the as-grown film. As the $Y^{\frac{1}{3}}$ method is only valid for nondegenerate semiconductors, we conclude that the $Y^{\frac{1}{2}}$ method for degenerate semiconductors is the most appropriate in the initial stages of ambient exposure. In fact, the $Y^{\frac{1}{2}}$ method yields an ionization energy of 5.72~eV on the as-grown surface, which is 20~meV larger than the work function. This value is in good agreement with (bulk) thermovoltage measurements, which yield a bulk doping density of $(2\pm1)\times 10^{18}$~cm$^{-3}$, thus implying a Fermi level ($2 \pm 27$)~meV above the valence band maximum (VBM) in the CuI bulk using Fermi statistics. A surface band diagram based on these values is sketched on the left side of Fig.~\ref{fig:humidity_model}(b), and labeled as "$ t = 0$" to represent the as-grown surface. Since the Fermi levels on the as-grown surface and in the bulk are found to coincide, the CuI bands and the vacuum level are drawn flat. On the other hand, applying the $Y^{\frac{1}{2}}$ method after 48~h of ambient exposure yields an ionization energy that is 360~meV larger than the work function (compare Fig.~\ref{fig:pes}(c) and Fig.~\ref{fig:time_dependent}(b)) significantly outside the applicability range of the method. Hence, the $Y^{\frac{1}{3}}$ method for nondegenerate semiconductors should be employed at this later stage. The corresponding ionization energy is 4.88~eV, which is 80~meV lower than the work function, thus confirming the applicability of the $Y^{\frac{1}{3}}$ method at this stage. A surface band diagram for the ambient-exposed CuI surface is shown on the right side of Fig.~\ref{fig:humidity_model}(b) and is labeled as "$t = 24$~h". The surface dipole, given by $\Delta \Phi = -0.90$~eV, is represented by the abrupt shift in the vacuum level $E_\mathrm{vac}$. Since the ionization energy decreases slightly less than the work function from $t = 0$ to $t = 24$~h ($-0.84$~eV versus $-0.90$~eV), we do not exclude the possibility of a small VBM downshift at the ambient-exposed surface due to chemical effects such as, e.g., surface oxidation.}

To investigate the origin of the conductivity enhancement upon water adsorption, the time-dependent conductivity experiment shown in Fig.~\ref{fig:time_dependent}(a) was repeated at \blue{six different RHs between 0 and 80\%}, with a fresh sample grown before each new experiment. At each RH, we plot the ratio between the maximum conductivity over time and the initial conductivity of the as-grown surface (Fig.~\ref{fig:various_humidities}(a)). This conductivity enhancement factor is maximized when RH is around 30-40\% and is negligible at very low and very high humidities. Additional CuI samples with a different thickness confirm this trend (Fig.~S6, Supporting Information). \blue{Simply plotting the maximum conductivity over time versus RH also results in a similar trend (Fig.~S7, Supporting Information).}
%We regard the conductivity enhancement factor as more robust metric than the maximum conductivity because the initial conductivity has a standard deviation around 20\% from sample to sample.
Since a RH of 30-40\% corresponds to about 1~ML coverage as explained above, we conclude that the conductivity-enhancing mechanism is maximized by the presence of no more than 1~ML of adsorbed water. Furthermore, the time evolution of conductivity and work function are \blue{similar to each other} (Figs.~\ref{fig:time_dependent}(a,b)) with a fast initial transient in the first couple of hours after exposure to moist air. These findings provide strong evidence against bulk effects, such as increased hole doping in CuI by incorporation of, e.g., O$_\mathrm{I}$ acceptors in the bulk. Instead, the conductivity enhancement seems to be related directly to the decrease in work function at CuI surfaces, although it cannot simply be attributed to a decrease in contact resistance as the four-point measurement configuration excludes contributions from contact resistance. Another option is the existence of a parallel conduction path through the adsorbed water layer \blue{at the film surface}, which could lead to an increase in the measured conductivity. However, the conductivity of adsorbed water is not higher than a few mS/cm
even under extreme humidity.~\cite{Guckenberger1994} As the conductivity of our as-grown CuI is about three orders of magnitude higher (Fig.~\ref{fig:time_dependent}(a)), the adsorbed water layer cannot be regarded as a high-conductivity path. \blue{Surface doping effects can also be excluded because the Fermi level of the humidity-exposed surface (right side of Fig.~\ref{fig:humidity_model}(b)) is further away from the VBM compared to the Fermi level of the as-grown surface (left side of Fig.~\ref{fig:humidity_model}(b)), indicating a lower surface hole concentration after water adsorption.}

\begin{figure}[t!]
\centering%
\includegraphics[width=0.8\columnwidth]{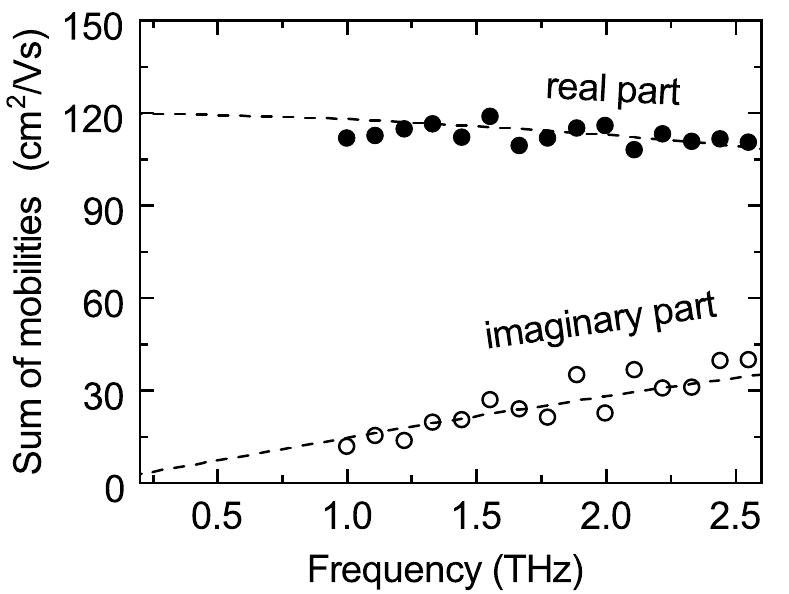}
\caption{Sum $\Sigma_\mu$ of electron- and hole mobility in a CuI film deposited on fused silica glass, as measured by THz \blue{transmission spectroscopy using a 400~nm laser pump to generate free carriers.} The dashed lines are fits to the real and imaginary parts of $\Sigma_\mu$ using the Drude model with a carrier effective mass of 0.3~$m_\mathrm{e}$ and a carrier scattering time of 20~fs. The extrapolated value of $\Sigma_\mu$ at \blue{zero frequency} (DC electric field) is 120~cm$^2$/Vs.
}
\label{fig:thz}
\end{figure}

%In the following paragraphs we propose a model which can qualitatively explain the moisture-enhanced conductivity consistent with existing electrical transport results in CuI films.~\cite{Kneiss2018} Note that we will refer to both the film surface and the grain boundaries as "surfaces", since grain boundaries are also subject to adsorption phenomena in polycrystalline materials.~\cite{Indekeu1991} It is likely that the surfaces of the as-grown films are preferentially iodine-terminated, due to the excess iodine vapor present during the conversion from Cu to CuI. The dangling bonds at iodine-terminated surfaces are known to be acceptor defects~\cite{Wang2012c} and are likely to be effective carrier scattering centers due to lack of passivation. Hence, we may attribute the degenerate doping observed on as-grown CuI surfaces (Fig.~\ref{fig:time_dependent}(b)) to those iodine dangling bonds. Degenerate surface doping could explain the weak 2D carrier antilocalization effect observed by low-temperature magnetoresistance.~\cite{Kneiss2018}

\blue{Instead, we propose the following explanation for the moisture-enhanced conductivity.
The as-grown surface is highly doped (Fig.~\ref{fig:humidity_model}(b)) but highly defective due to unpassivated surface dangling bonds. Once a ML of water has been adsorbed, the surface becomes more weakly doped (Fig.~\ref{fig:humidity_model}(b)) but the surface dangling bonds become saturated by water molecules. Since the grain boundaries of various polycrystalline iodides are known to experience fast hydration in a humid environment~\cite{Leguy2015,BradleyPhipps1981} we assume that the CuI grain boundaries become hydrated in a similar time frame as the film surface characterized in this work. The phenomena discussed above are unlikely to enhance the hole conductivity \textit{along} the film surface or grain boundaries due to the lower hole concentration of the humidity-exposed surface. However, it is reasonable to expect enhanced hole transport \textit{across} grain boundaries because of water passivation of dangling bonds at grain boundaries. Furthermore, the presence of a conductive water layer between grain boundaries may open up more physical transport channels across poorly connected grain boundaries. Despite the presence of a hole barrier due to the water-induced surface dipole (Fig.~\ref{fig:humidity_model}(b)), the thickness of a water ML is only $\sim$ 1-3~\AA, so the tunneling probability for holes across grain boundaries is high.
%Together with the decrease in work function, water adsorption on CuI leads to saturation of surface dangling bonds, possibly in conjunction with desorption of the volatile iodine atoms at the surface. Hence, we speculate that water adsorption changes the nature of hole transport at the film surface. Specifically, the as-grown surface is highly doped and highly defective to to lack of passivation.
%In fact, the as-grown surface is likely to be iodine-terminated due to the excess iodine vapor present during film growth, and dangling bonds at iodine-terminated surfaces are known to form acceptor defects.~\cite{Wang2012c}
%Once a ML of water has been adsorbed, the surface becomes more weakly doped (lower hole concentration) but the 
%, to less defective hole barriers once a monolayer of water has formed.} 
If this interpretation of humidity-enhanced conductivity is correct, then grain boundaries act as bottlenecks for hole transport in our CuI films. Thus, one would expect the intra-grain hole mobility in CuI to be significantly larger than the long-range hole mobility including transport across various grain boundaries. The former can be measured by THz spectroscopy~\cite{Hempel2016} and the latter can be measured using the Hall effect. Indeed, our THz spectroscopy measurements on CuI indicate Drude-like conductivity and a value of 120~cm$^2$/Vs as the sum of carrier mobilities (Fig.~\ref{fig:thz}). This value is much larger than the hole mobility found by Hall measurements ($\sim 6$~cm$^2$/Vs) for films of similar conductivity synthesized in the same way.~\cite{Schein2013} This discrepancy is compatible with the hypothesis that grain boundaries limit charge transport in CuI films. This limitation can be mitigated by water adsorption and grain boundary hydration. Interestingly, previous magnetoresistance experiments also attributed the high mobilities of CuI films to tunneling effects across grain boundaries, although the effect of grain boundary hydration was not considered.~\cite{Kneiss2018}}
There are no reasons to expect that passivation of CuI \blue{grain boundaries} should improve after the first monolayer of water has been adsorbed, so the conductivity enhancement factor is not expected to increase at higher humidities, where more than one stable monolayer of water exists. In fact, the conductivity enhancement factor even drops at higher humidities (Fig.~\ref{fig:various_humidities}(a)) due to other unidentified loss mechanisms, such as water absorption in the bulk, or to a too large surface dipole.

Finally, we show in Fig.~\ref{fig:various_humidities}(b) that humidity-dependent conductivity changes have both a reversible and an irreversible component. Unlike the case of Fig.~\ref{fig:various_humidities}(a), in this experiment the \textit{same} sample was exposed to different humidities sequentially, and its conductivity was monitored. At each RH change, the measured conductivity changes consistently with the data in Fig.~\ref{fig:various_humidities}(a). For example, the conductivity increases when moving from 17\% to 60\% RH or from 60\% to 35\% RH, and it decreases when moving from 35\% to 20\% RH or from 20\% to 10\% RH (Fig.~\ref{fig:various_humidities}(b)). However, there is also a long-term overall trend of decreasing conductivity, as evident when comparing the conductivity at 35\% after $\sim$50~hr and the conductivity at 30-40\% after $\sim$120~hr. Slow iodine re-evaporation from the film could be responsible for this effect, since we also observe a slow decrease in conductivity in a CuI film stored in a water-free atmosphere in nitrogen (Fig.~\ref{fig:time_dependent}(a)). The occurrence of iodine re-evaporation was confirmed by placing a CuI film next to a Cu plate in a small sample box and observing coloration of the Cu plate after a few days of storage in air.

\section{Conclusion}
CuI films exposed to moist air experience a conductivity increase and a work function decrease over a similar time frame (a couple of hours). We attribute the conductivity increase to mitigation of grain boundary losses by hole tunneling through water-passivated grain boundaries. The conductivity increase is partially reversible and is maximized at 30-40\% RH, which probably corresponds to one monolayer of adsorbed water. We encourage other researchers to report the atmosphere exposure time and humidity level when presenting CuI characterization results.

%\vspace{0.5cm}
%\section*{Supporting information}

\section*{Supplementary information}
See supplementary material for structural charaterization by XRD and Raman spectroscopy, band gap determination by optical spectroscopy, additional humidity-dependent conductivity experiments, and more detailed plots of the initial and final phases of the conductivity, work function, and ionization \blue{energy} measurements.

%\vspace{0.5cm}
\section*{Acknowledgements}
This project has received funding from the European Union’s Horizon 2020 research and innovation programme under the Marie Sk\l odowska-Curie grant agreement No 840751. We acknowledge Lars Steinkopf, Sergiu Levcenko, and Jos\'e M\'arquez Prieto for technical assistance and helpful discussions.

%\section*{References}
\bibliography{library}

\end{document}